\documentclass[aps,showpacs,preprint,superscriptaddress]{revtex4}
\usepackage{graphicx}
\usepackage{subfigure}
\usepackage{float}
\usepackage{amsmath}
\usepackage{amsfonts}
\usepackage{bm}
\usepackage{bbm}
\usepackage{txfonts}
\usepackage{array}
\usepackage{amssymb}

\begin{document}
\title{Enhanced pair production in multi-pulse trains electric fields with oscillation}
\author{Lie-Juan Li}
\affiliation{School of Mathematics and Physics, Lanzhou Jiaotong University, Lanzhou 730070, China}
\affiliation{Key Laboratory of Beam Technology of the Ministry of Education, and College of Nuclear Science and Technology, Beijing Normal University, Beijing 100875, China}
\author{Xiao-Wei Sun}
\affiliation{School of Mathematics and Physics, Lanzhou Jiaotong University, Lanzhou 730070, China}
\author{Melike Mohamedsedik}
\affiliation{Key Laboratory of Beam Technology of the Ministry of Education, and College of Nuclear Science and Technology, Beijing Normal University, Beijing 100875, China}
\affiliation{College of Xinjiang Uyghur Medicine, Hotan 848000, China}
\author{Li Wang}
\affiliation{Institute of Radiation Technology, Beijing Academy of Science and Technology, Beijing 100875, China}
\author{Li-Na Hu}
\affiliation{Key Laboratory of Beam Technology of the Ministry of Education, and College of Nuclear Science and Technology, Beijing Normal University, Beijing 100875, China}
\author{Bai-Song Xie \footnote{bsxie@bnu.edu.cn}}
\affiliation{Key Laboratory of Beam Technology of the Ministry of Education, and College of Nuclear Science and Technology, Beijing Normal University, Beijing 100875, China}
\affiliation{Institute of Radiation Technology, Beijing Academy of Science and Technology, Beijing 100875, China}
\date{\today}

\begin{abstract}
For different alternating-sign multi-pulse trains electric fields with oscillation, the effects of the electric field pulse number and the relative phase of the combined electric field on pair production are investigated by solving quantum Vlasov equation.
It is found that the number density of created particles in the combined electric fields is increased by more than one order of magnitude compared to the results without oscillating structure for both zero transverse momentum and full momentum space.
In the case of zero transverse momentum, the created particles longitudinal momentum spectrum are monochromatic for large pulse numbers and some suitable relative phases. The number density depends nonlinearly on the relative phase that enables the optimal relative phase parameters for the number density. Moreover, for the full momentum space, the created particles number density and momentum spectrum under different multi-pulse trains electric fields are given and discussed. We also find that the number density as a function of pulse number satisfies the power law with index $5.342$ for the strong but slowly varying electric field with large pulse numbers.



\end{abstract}
\pacs{12.20.Ds, 03.65.Pm, 02.60.-x}
\maketitle

\section{Introduction}
Nonperturbative electron-positron pair production
from vacuum in the presence of strong external fields is a
nontrivial prediction of quantum electrodynamics (QED)\cite{Sauter:1931zz,Dirac:1928ej,Anderson:1933mb,Dunne:2008kc,Heisenberg:1936nmg}.
It is yet to be observed due to its strong exponential suppression of the order of
$\exp (-\pi E_{cr}/E)$, where
$E_{cr} =m_{e} ^{2} c ^{3} / e \hbar \simeq 1.3 \times {10 ^{16}} \rm V/ \rm cm$
is the Schwinger critical field strength and $E$ is the constant field strength\cite{Schwinger:1951nm}.
Obviously, the corresponding laser intensity is $ I_{cr} \simeq 10^{29} \rm W / \rm cm^{2}$
and it is still much higher than the current laser intensity.
In the perturbation domain, multi-photon pair production as an another important mechanism has already been experimentally
verified at the pioneering SLAC E-$144$ experiment\cite{SLAC E-144:1997}.
Many recent theoretical investigation results propose that the pair creation
rate can be tremendously increased by careful shaping the laser pulse field \cite{Dunne:2009gi, DiPiazza:2009py, Bulanov:2010ei, Bell:2008zzb} and/or  utilizing both mechanism together\cite{Schutzhold:2008pz, Orthaber:2011cm, Olugh:2019nej,Li:2021vjf}, which brings hope for the upcoming experiments with advanced laser
facilities\cite{Ringwald:2001ib,Marklund:2008gj,Heinzl:2009bmy,Olugh:2019nej,Pike:2014wha}

More importantly, many theoretical methods are presented to study the creation of particle pairs
under different external field configurations in order to reduce the pair production threshold or/and increase the
total number of created
particles\cite{Hebenstreit:2009km,Hebenstreit:2010,Dumlu:2010vv,Hebenstreit:2011wk, Kohlfurst:2017git,Wang:2019oyk,Ababekri:2019dkl,Kohlfurst:2019mag, Mohamedsedik:2021pzb,Fedotov:2022ely,Xie:re2017}. There are two commonly
used theoretical methods to investigate it. One is the
semiclassical approximation method, such as the generalized
Wentzel-Kramers-Brillouin (WKB) approximation\cite{Brezin:1970xf,Kim:2007pm,Kim:2003qp}
and the worldline instanton technology\cite{Kim:2000un,XieCPL}.
The other is the quantum kinetic method, such as the
quantum Vlasov equation (QVE)
\cite{Kchmidt:1998,Bloch:1999,Hebenstreit:2009km,Kohlfurst:2012rb,Abdukerim:2013vsa,Nuriman:2012hn}
, the low density approximation\cite{Blaschke:2012vf} and
Dirac-Heisenberg-Wigner (DHW) approach\cite{1991:Phase-space,Hebenstreit:2010,Hebenstreit:2011wk}.
For instance, Brezin and Itzykson
studied the generation of pairs under spatially homogeneous time-varying electric field by WKB approximation\cite{Brezin:1970xf} and gave
the particle pairs creation probability in different regions of $\gamma \ll1$ and $\gamma \gg1$, where Keldysh adiabaticity parameter $\gamma$ depends on the external field frequency $\omega$ and strength $E$, i.e., $\gamma=m\omega/eE$ ($m$ is the particle mass, and $e$ is the particle charge). Hebenstreit et al.\cite{Hebenstreit:2009km} studied the
particle pairs creation in short laser pulses with subcycle structure by employing quantum kinetic approach and found that the momentum spectrum is
very sensitive to the parameters of the applied field. Sch\"utzhold
et al.\cite{Schutzhold:2008pz} proposed that the dynamically assisted Schwinger mechanism can lead to a significant enhancement of the tunneling process by superimposing a
strong but slowly varying electric field on a weak but rapidly
changing one.

Moreover, it is also well known that time-domain multiple-slit interference effect
as a new route to enhance the pair production
has been widely explored, because it can not only
obtain some nontrivial phenomena of the momentum spectra, but also strongly increase the pair creation rate.
Specifically, Akkermans and Dunne\cite{Akkermans:2011yn} proposed the Ramsey multiple-time-slit interference effect and
found that the central value of momentum spectrum for an
alternating-sign $N$-pulse electric field grows like $N^{2}$.
Kohlf\"urst \cite{Kohlfurst:2012yw} explored the interference effects arising in various field configurations with multiple pulses
and pointed out that the particle number density can be optimized by pulse-shaping.
Li et al.\cite{Li:2014xga} investigated the relationship between the number density
of created bosons $n$ and the electric field pulse number $N$ in an alternating-sign
$N$-pulse electric field. The results show that the number density of created
bosons increases with the pulse number for zero transverse momentum.
More recently, they also studied the pair production
in strong fields by multiple-slit interference effect
with dynamically assisted Schwinger mechanism, and proposed that
the relationship between the number density of created particles
and the electric field pulse number satisfies a linear relationship for the
full momentum space\cite{ Li:2014psw} .
To our knowledge, the frequency plays an important role for vacuum pair production in strong field QED and the types of fields represent more realistic pulse configurations with rich behavior in the momentum spectrum of the created particles
\cite{Hebenstreit:2009km,Ruf:2008ahs,Dumlu:2010ua,Dumlu:2010vv,Dumlu:2011rr}.
In addition, the relative phase has a significant
effect on the electron-positron pair creation process in the spatially inhomogeneous combined field\cite{Li:2021wag}.

In this work, we will entirely focus on the quantum kinetic approach
to study the pair production in several alternating-sign multi-pulse trains electric fields with oscillation. The background field that we considered including a strong but slowly varying one, a weak but rapidly changing one as well as the combined one.
The influences of the electric field pulse number $N$ and the
relative phase $\varphi$ on the momentum spectrum and the number density of created particles are studied, and the possible explanation for obtained results are also given and discussed.
One expect that the exploration is helpful to obtain more higher particle number by pulse-shaping and valuable to observe
electron-positron in nonperturbative regime for the coming experiments.
\section{THEORETICAL FRAMEWORK}\label{theorecical framwork}

\subsection{Oscillating multi-pulse trains electric fields}\label{field}

For our pair production studies, we consider the alternating-sign multi-pulse trains electric field with oscillatory structure of the following form
\begin{equation}\label{FieldMode}
\begin{aligned}
E\left(t\right)
&= E_{1s}\left(t\right)+E_{2w}\left(t\right)\\
&=\sum_{i=0}^{N-1}(-1)^i E_{1s} \text{sech}^{2} \left[ \frac{t-i T_{D}}{\tau_{1}}\right]\cos(\omega_{1}(t-i T_{D})) \\
&+\sum_{i=0}^{N-1}(-1)^i E_{2w} \text{sech}^{2} \left[ \frac{t-i T_{D}}{\tau_{2}}\right]\cos(\omega_{2}(t-i T_{D})+\varphi),
\end{aligned}
\end{equation}
where $E_{1s}\left(t\right)$ and $E_{2w}\left(t\right)$ denote the electric field pulse trains with alternating sign of the low-frequency
strong one and the high-frequency weak one,
$N$ represents the electric field pulse number, $E_{1s,2w}$ are the strength of the background electric fields, $\tau_{1,2}$ are the time width scales,
$\omega_{1,2}$ indicate the field oscillating frequencies, $T_{D}$ sets the time interval between the electric field pulses
and $\varphi$ is relative phase between the strong and weak electric fields.
For this type of electric field configuration, we use the temporal gauge $A_{0}(t)=0$, and the corresponding vector potential
becomes ${\mathbf A_{\mu}}(t)=(0,0,0,A(t))$ with the field strength $E(t)=-\dot{A}(t)$.
In this investigation we choose the values of $E_{1s,2w}$, $\tau_{1,2}$ and $T_{D}$ applied in previous
work \cite{Li:2014psw} and also add other parameters such as $\omega_{1,2}$ and $\varphi$. Thus, the studied parameter values
are chosen as $E_{1s}=0.1 E_{cr}$, $\tau_{1}=1/0.02$ , and $\omega_{1}=0.05$ for $E_{1s}\left(t\right)$.
For $E_{2w}\left(t\right)$, the fixed parameter values are chosen as  $E_{2w}=0.01 E_{cr}$,
$\tau_{2}=1/0.22$, $\omega_{2}=0.5$. The time interval is $T_{D}=400.3$. Throughout this paper, we use natural unites $\hbar=c=1$ and set the electron mass $m=1$.

\subsection{Theoretical description: Quantum Vlasov equation}\label{method}

To study pair creation in the above mentioned scenario, we employ the QVE, which is one of the quantum
kinetic theory and has been widely used to investigate pair production under time dependent fields with spatial homogeneity \cite{Nuriman:2012hn,Kohlfurst:2012rb,Abdukerim:2013vsa}.
As the specific derivation of QVE has been given in previous works \cite{Kchmidt:1998,Bloch:1999}, we only present the basic
ideas and the essential points of this method.
Beginning with the Dirac equation in a homogeneous electric field and applying the a canonical time-dependent
Bogoliubov transforation, we can obtain QVE from the following integro-differential equation about the
one-particle momentum distribution function $f({\mathbf k},t)$
\begin{equation}\label{intediffEquation}
\frac{df(\mathbf k,t)}{dt} = \frac{1}{2}\frac{eE(t)\varepsilon_{\perp}}{\Omega^{2}(\mathbf{k},t)}
\int_{t_{0}}^{t}dt^{'}\frac{eE(t^{'})\varepsilon_{\perp}}{\Omega^{2}(\mathbf{k},t^{'})}
[1-2f({\mathbf k},t^{'})]\times\cos[2\int_{t^{'}}^{t}d\tau\Omega(\mathbf k,t)],
\end{equation}
where $e$ is the electron charge, $\mathbf k=(\mathbf k_{\perp},k_{\parallel})$ is the canonical momentum,
$\Omega(\mathbf{k},t)=\sqrt{\varepsilon_{\perp}^{2}+p_{\parallel}^{2}(t)}$ represents the total energy,
$\varepsilon_{\perp}=\sqrt{m^{2}+\mathbf k_{\perp}^{2}}$ is the transverse energy, $m$ is the electron mass,
$p_{\parallel}(t)=k_{\parallel}-eA(t)$ denotes the kinetic momentum is related to the longitudinal canonical momentum
$k_{\parallel}$ and the direction along the electric field $E(t)$. Note that the one-particle
momentum distribution function $f({\mathbf k},t)$ describes the created real particles at
 $t\rightarrow +\infty$,
where the applied electric field will become zero because of the practical external
laser field is turned on and turned off at the finite time.
Thus, we focus our attention to the investigation of
distribution function $f({\mathbf k},+\infty)$ and the corresponding particle number density $n(+\infty)$.

We define $W(\mathbf k,t)= eE(t)\varepsilon_{\perp}/\Omega^{2}(\mathbf{k},t)$ and
$\Theta(\mathbf k,t^{'},t) =\int_{t^{'}}^{t}d\tau\Omega(\mathbf k,t) $ then
Eq.(\ref{intediffEquation}) becomes
\begin{equation}\label{reducedEq}
\frac{df(\mathbf k,t)}{dt} = \frac{1}{2}W(\mathbf k,t)
\int_{t_{0}}^{t}dt^{'}W(\mathbf k,t)
[1-2f({\mathbf k},t^{'})]\times\cos[2\Theta(\mathbf k,t^{'},t)],
\end{equation}
For computational reasons, two quantities are introduced as
\begin{equation}\label{Quantities1}
 u(\mathbf k,t)=\int_{t_{0}}^{t}dt^{'}W(\mathbf k,t)
[1-2f({\mathbf k},t^{'})]\times\cos[2\Theta(\mathbf k,t^{'},t)],
\end{equation}
\begin{equation}\label{Quantities2}
 v(\mathbf k,t)=\int_{t_{0}}^{t}dt^{'}W(\mathbf k,t)
[1-2f({\mathbf k},t^{'})]\times\sin[2\Theta(\mathbf k,t^{'},t)],
\end{equation}
Eq.(\ref{reducedEq}) can be rewritten as a first order, ordinary differential equation(ODE) system
\begin{equation}\label{ODE1}
 \frac{df(\mathbf k,t)}{dt}=\frac{1}{2}W(\mathbf k,t)u(\mathbf k,t)
\end{equation}
\begin{equation}\label{ODE2}
 \frac{du(\mathbf k,t)}{dt}=W(\mathbf k,t)[1-2f({\mathbf k},t)]-2\Omega(\mathbf k,t)v(\mathbf k,t)
\end{equation}
\begin{equation}\label{ODE3}
 \frac{dv(\mathbf k,t)}{dt}=2\Omega(\mathbf k,t)u(\mathbf k,t)
\end{equation}
Together with the initial conditions
$f(\mathbf{k},-\infty)=u(\mathbf{k},-\infty)=
v(\mathbf{k},-\infty)=0$
this set of equations become a well-defined and numerically straightforward solvable initial
value problem.
The number density of created pairs is obtained by integrating the distribution function
$f(\mathbf{k},t)$ to all momenta at asymptotically value $t\to +\infty$
\begin{equation}\label{3}
  n = 2\lim_{t\to +\infty}\int\frac{d^{3}q}{(2\pi)^ 3}f(\mathbf{k},t) \, .
\end{equation}

\section{NUMERICAL RESULTS}\label{results}

In this section, we give the main numerical results and the estimates of the created particles momentum spectra as well as number density for the case of above electric fields models.
Moreover, we know that there is a tight connection of the momentum distribution and number density, thus the particle number density $n(\mathbf{k}_{\perp}=0)$ changes with
relative phase $\varphi$ for the combined electric field $E(t)$
with different pulse number $N$ is further investigated.

\subsection{Results for the vanishing transverse momentum space}

\begin{figure}[ht]\suppressfloats
\includegraphics[scale=0.5]{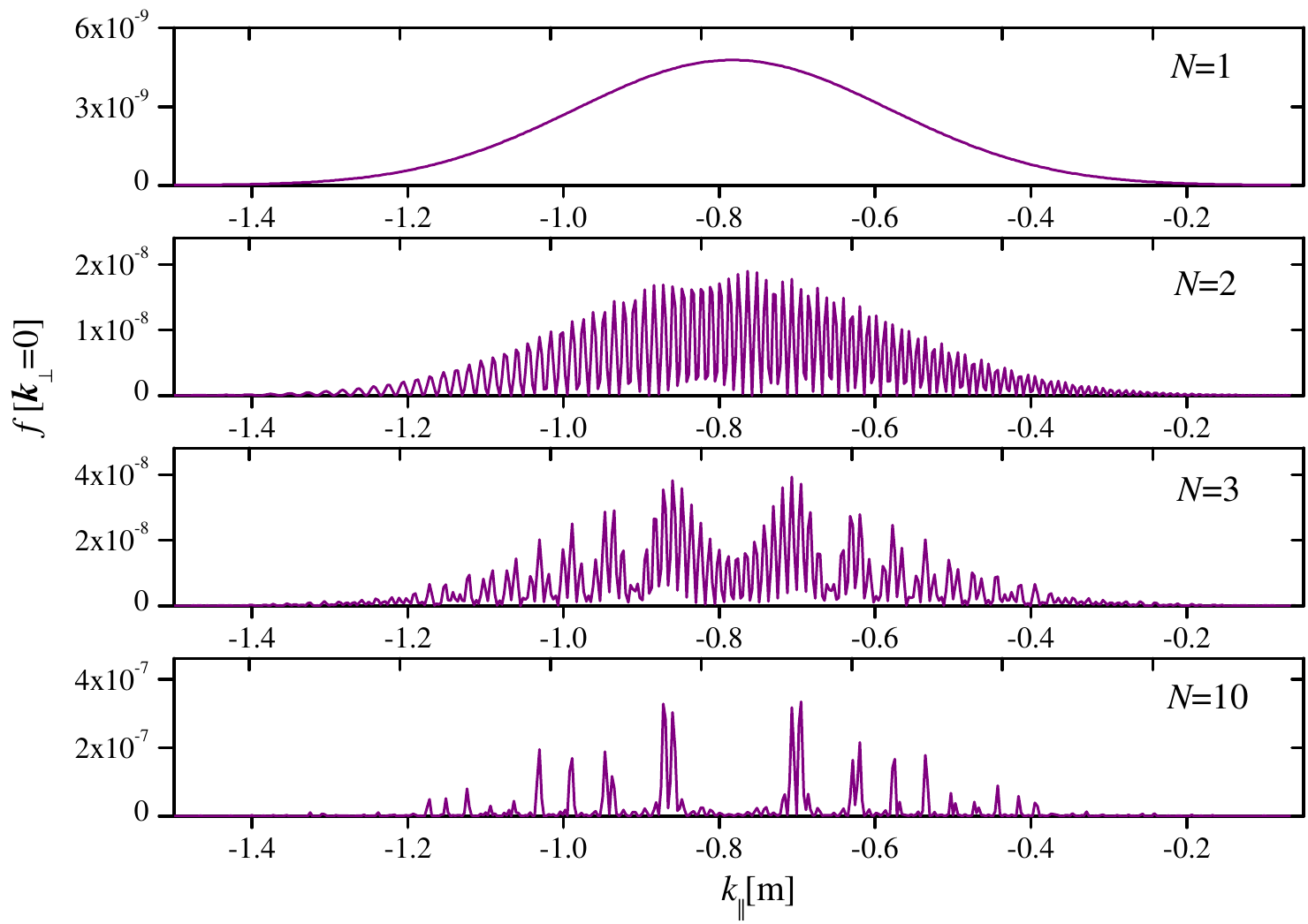}
\caption{(color online). The longitudinal momentum spectra of
created particles in the combined electric field
$E(t)=E_{1s}(t)+E_{2w}(t)$ with
various electric-field pulse numbers
of $N$ as $1$, $2$, $3$ and $10$. The electric field parameters are
$E_{1s}=0.1 E_{cr}$, $\tau_{1}=1/0.02$, $\omega_{1}=0.05$,
$E_{2w}=0.01 E_{cr}$,
$\tau_{2}=1/0.22$, $\omega_{2}=0.5$, $T_{D}=400.3$ and $\varphi=0$.}
\label{fig:1}
\end{figure}

Let us consider first the alternating-sign multi-pulse trains electric fields with vanishing transverse momentum, i.e., $\mathbf{k}_{\perp}=0$. The longitudinal momentum spectra of
created particles in the combined electric field
$E(t)=E_{1s}(t)+E_{2w}(t)$ with
various electric-field pulse numbers are shown in Fig.\ref{fig:1}, where the pulse
number is chosen as $1$, $2$, $3$ and $10$. It can be seen
that for the
single-pulse electric field, i.e., $N=1$, the longitudinal momentum spectrum shows a Gaussian-like shape and there is no oscillatory behavior.
For the larger pulse number ($N\geq 2$), we see the formation of oscillatory pattern in the
longitudinal momentum spectrum.
The oscillations can be understood from the viewpoint of the scattering potential because of the interference effect between separate complex conjugate pairs of turning points.
From Fig.\ref{fig:1}, we can see that the longitudinal momentum spectrum for the odd-pulse electric field ($N=1,3$) are symmetric and asymmetric for the even-pulse electric field ($N=2,10$).
This result is related to the electric field symmetry,
when the pulse number $N$ is odd, the electric field proposed in
Eq.\ref{FieldMode} is an even function of time and the corresponding vector potential is an odd function of time, which affect the symmetry of the total energy $\Omega(\mathbf{k},t)$ and further determine the symmetry of distribution function $f({\mathbf k},+\infty)$.

Additionally, it can be clearly seen that
the peak value of the momentum spectrum increases significantly
and momentum distribution range shrinks
with the increase of electric field pulse number.
The relationship between the central momentum distribution function
and the pulse number $N$ is complex and different for various electric fields.
In the case of $N \geq 2$, there is no a certain relationship between the central momentum distribution function
and the pulse number.
For the large pulse number $N=10$, a quasi-single peak
appears in the momentum spectrum at the specific momentum value.
As a result, the created particles spectrum become more monoenergetic
for the large pulse number and the quasi-monoenergetic particles can be
obtained by optimizing pulse number.

\begin{figure}[htbp]\suppressfloats
\includegraphics[scale=0.5]{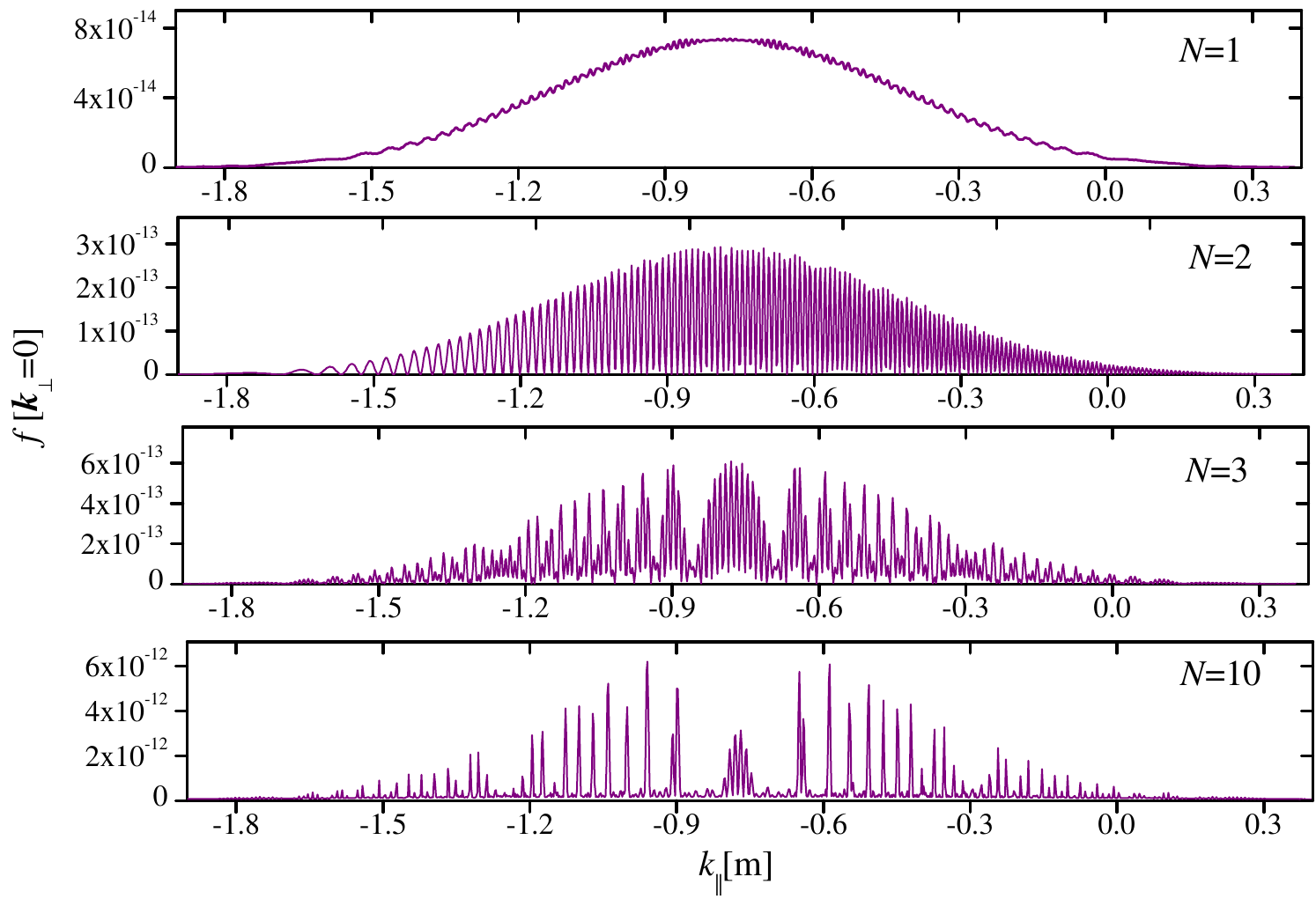}
\caption{(color online). The longitudinal momentum spectra of
created particles in the low-frequency strong field pulse train $E_{1s}(t)$ with various electric field pulse numbers of $N$ as $1$, $2$, $3$ and $10$. The electric field parameters are $E_{1s}=0.1 E_{cr}$, $\tau_{1}=1/0.02$, $\omega_{1}=0.05$ and $T_{D}=400.3$.}
\label{fig:2}
\end{figure}
To further investigate the effect of oscillation on the electron-positron
pair production, we now consider a strong but slowly varying
electric field pulse train.
The longitudinal momentum spectra of
created particles for different pulse numbers $N$ are sketched in
Fig. \ref{fig:2}. It is clear to see that there is
oscillation structure in momentum spectrum for the single-pulse electric field, i.e., pulse number $N=1$. In order to understand this behavior, we make qualitatively analysis by using WKB method \cite{Dumlu:2010ua} and find that there are two pairs turning points have a comparable distance from the real axis in the turning points structure. Note that in the semiclassical regime, the number density is depends on the turning points that are nearest to the real axis, while the degree of interference effect is dominated by the pair number of these turning points  that possess a similar distance to the real axis \cite{Dumlu:2011rr}. According, in the case of $N=1$, the momentum spectrum exhibits weak interference effect is understandable.
From other panels of Fig.\ref{fig:2}, one can see that for
$N=2$, $3$ and $10$, more obvious interference effect appear in the
longitudinal momentum spectra as compared to $N=1$.
By comparing with the single-pulse field of $N=1$
shown in top panel of Fig.\ref{fig:2}, we also find that, in the
case of the larger pulse number($N\geq 2$), the maximum value of the
distribution function becomes larger with increasing $N$.
Specifically, the maximum value of the
distribution function increases from $7.38\times 10^{-14}$ for $N=1$ to
$6.39\times 10^{-12}$ for $N=10$ enhanced almost $2$ orders of magnitude.

\begin{figure}[htbp]\suppressfloats
\includegraphics[scale=0.5]{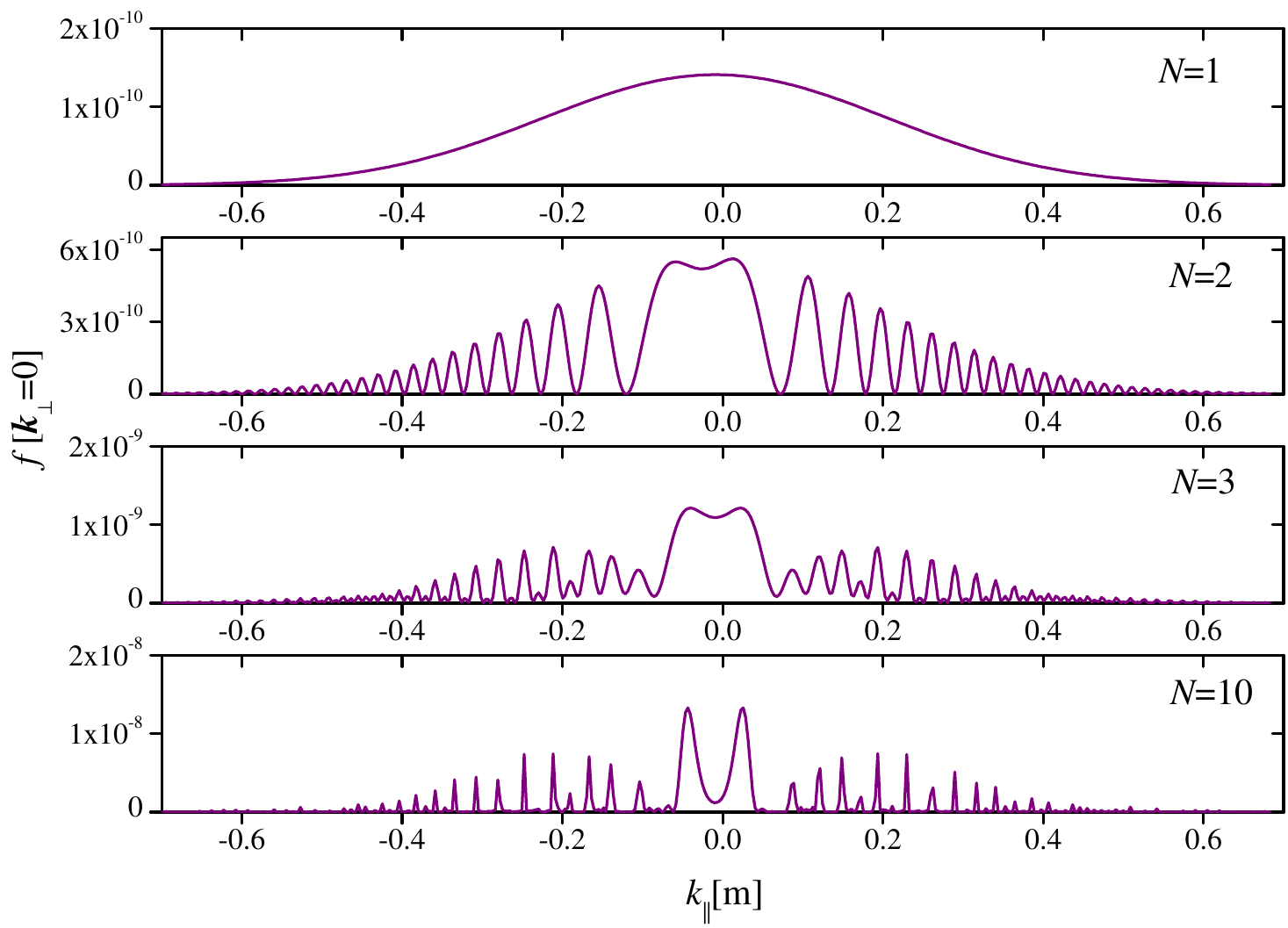}
\caption{(color online). The longitudinal momentum spectra of
created particles in the high-frequency
weak field pulse train $E_{2w}(t)$ with various electric-field
pulse numbers of $N$ as $1$, $2$, $3$ and $10$. The electric field parameters are $E_{2w}=0.01 E_{cr}$,
$\tau_{2}=1/0.22$, $\omega_{2}=0.5$, $T_{D}=400.3$ and $\varphi=0$.}
\label{fig:3}
\end{figure}

For the weak but rapidly changing electric field pulse train, the longitudinal momentum spectra of created particles for various electric-field
pulse numbers are shown in Fig.\ref{fig:3}.
In the case of single-pulse electric field, we see that the
longitudinal momentum spectrum has no interference.
However, for $N \textgreater 1$, it can be seen that the momentum
spectra in the external field have a remarkable interference
pattern and the interference effect becomes more and more
obvious with the increase of pulse number.
This result can be understood qualitatively by
the structure of turning points.
Also, the peak value of longitudinal momentum spectrum
increases strongly with increasing the pulse number.
Note that the symmetry and monoenergetic behavior of the longitudinal momentum spectrum that mentioned for the combined electric field can also be observed from that for the high-frequency electric field pulse train.

\begin{figure}[htbp]\suppressfloats
\includegraphics[scale=0.5]{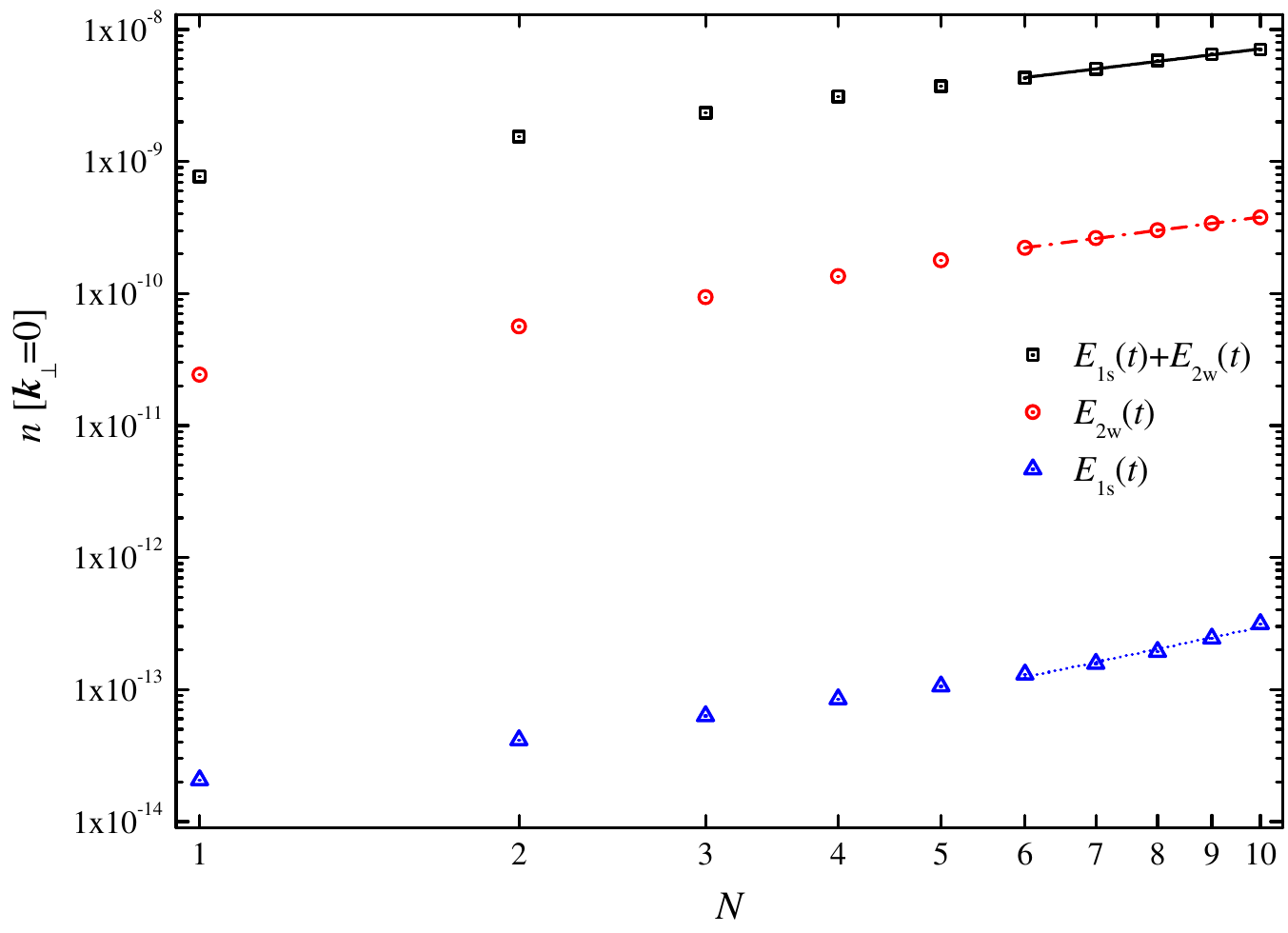}
\caption{(color online). Particles number density
$n(\mathbf{k}_{\perp}=0)$ changes with
electric field pulse number $N$ for the combined electric field $E(t)$ (black squares), the electric field $E_{2w}(t)$ (red circles), and the electric field $E_{1s}(t)$ (blue triangles), respectively. The black solid , red dash dotted and
blue dotted lines are the fitted ones for $E(t)$, $E_{2w}(t)$ and $E_{1s}(t)$
with the slopes approximate to $0.977$,$1.046$ and $1.714$. The electric
field parameters are consistent with those in Fig. \ref{fig:1}.}
\label{fig:4}
\end{figure}

In order to investigate the dependence of number density on the
electric field pulse number, we change pulse number $N$ from $1$ to $10$. The results for the different electric fields are depicted in Fig.\ref{fig:4}.
It is found that the created particles number density $n(\mathbf{k}_{\perp}=0)$ increases greatly with
increasing pulse number $N$.
In Fig.\ref{fig:4}, the solid black, red dash dotted and
blue dotted lines are the fitted ones for $E(t)$, $E_{2w}(t)$ and $E_{1s}(t)$.
We can see that the relationship between the number density and large pulse
number ($N \geq 6$) are approximately the power laws for $E(t)$, $E_{2w}(t)$ and $E_{1s}(t)$ and the different index as $0.977$, $1.046$ and $1.714$, respectively.
Furthermore, it is easy to see that for a certain number pulse,
the number density of created particles in the combined electric field is significantly increased compared to the results under a single electric field, and it is much higher than that obtained by simply adding of the single field. These result indicate that the coupling between the multiple-slit interference effect and dynamically assisted Schwinger mechanism leads to a strong enhancement of the electron-positron pair production.

\begin{figure}[htbp]\suppressfloats
\includegraphics[scale=0.5]{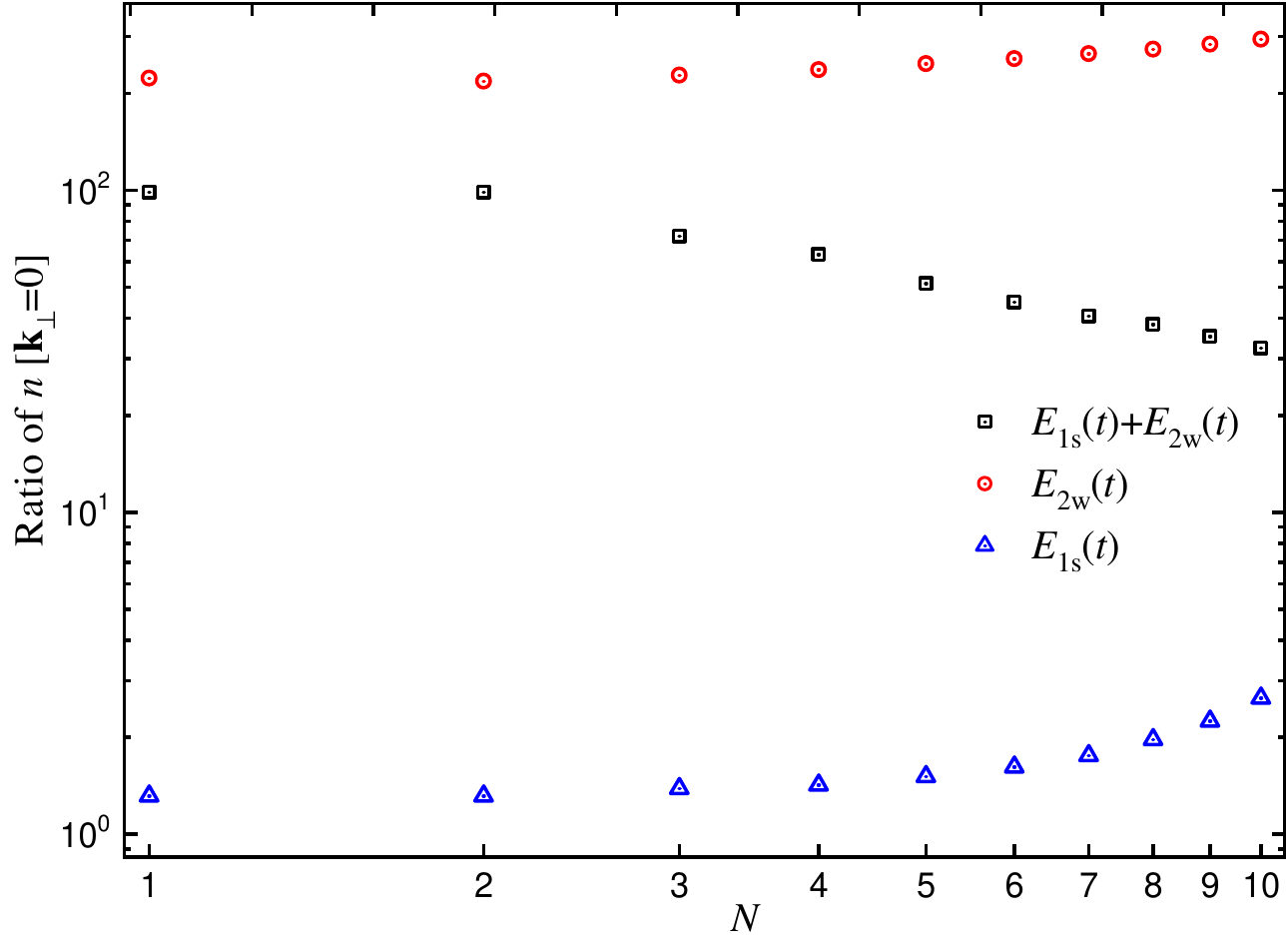}
\caption{(color online). Ratio of the created particle number density changes with
electric pulse number $N$, i.e., $n_{osc}(\mathbf{k}_{\perp}=0)/n_{nosc}(\mathbf{k}_{\perp}=0)$, for the combined electric field $E(t)$ (black squares), the high-frequency weak field $E_{2w}(t)$ (red circles) and the low-frequency strong field
$E_{1s}(t)$ (blue triangles) with oscillation in present study and without oscillation in Ref.\cite{Li:2014psw} when $\varphi=0$. The electric field parameters are consistent with those in Fig. \ref{fig:1}.}
\label{fig:5}
\end{figure}

More importantly, we explore the ratio of the particles number density of oscillation in present study and without oscillation in Ref.\cite{Li:2014psw} changes with the electric pulse number for different
electric fields when $\varphi=0$ and define it
as $n_{osc}(\mathbf{k}_{\perp}=0)/n_{nosc}(\mathbf{k}_{\perp}=0)$. The result
is plotted in Fig.\ref{fig:5} for the combined electric field $E(t)$ (black squares), the high-frequency weak field $E_{2w}(t)$ (red circles) and the low-frequency strong field
$E_{1s}(t)$ (blue triangles) when $\varphi=0$. From Fig.\ref{fig:5}, one can see
that the ratio value is always greater than $1$ when the pulse number changes.
It means that when the pulse number is fixed, the number density under the external field with oscillating structure is always higher than the results without it.
For the electric field pulse train $E_{2w}(t)$, the ratio
first decreases slightly and then increases with the increases of pulse number $N$. For the case of $E_{1s}(t)$, the ratio increases with the increases of $N$. We also obtain that
the ratio reaches maximum at the pulse number $N=2$ for the combined electric field.
Combining the above results of Fig.\ref{fig:5}, we find that the pairs creation rate can be tremendously increased by more than two orders in $E_{2w}(t)$ when the oscillation was considered.

\begin{figure}[htbp]\suppressfloats
\includegraphics[scale=0.5]{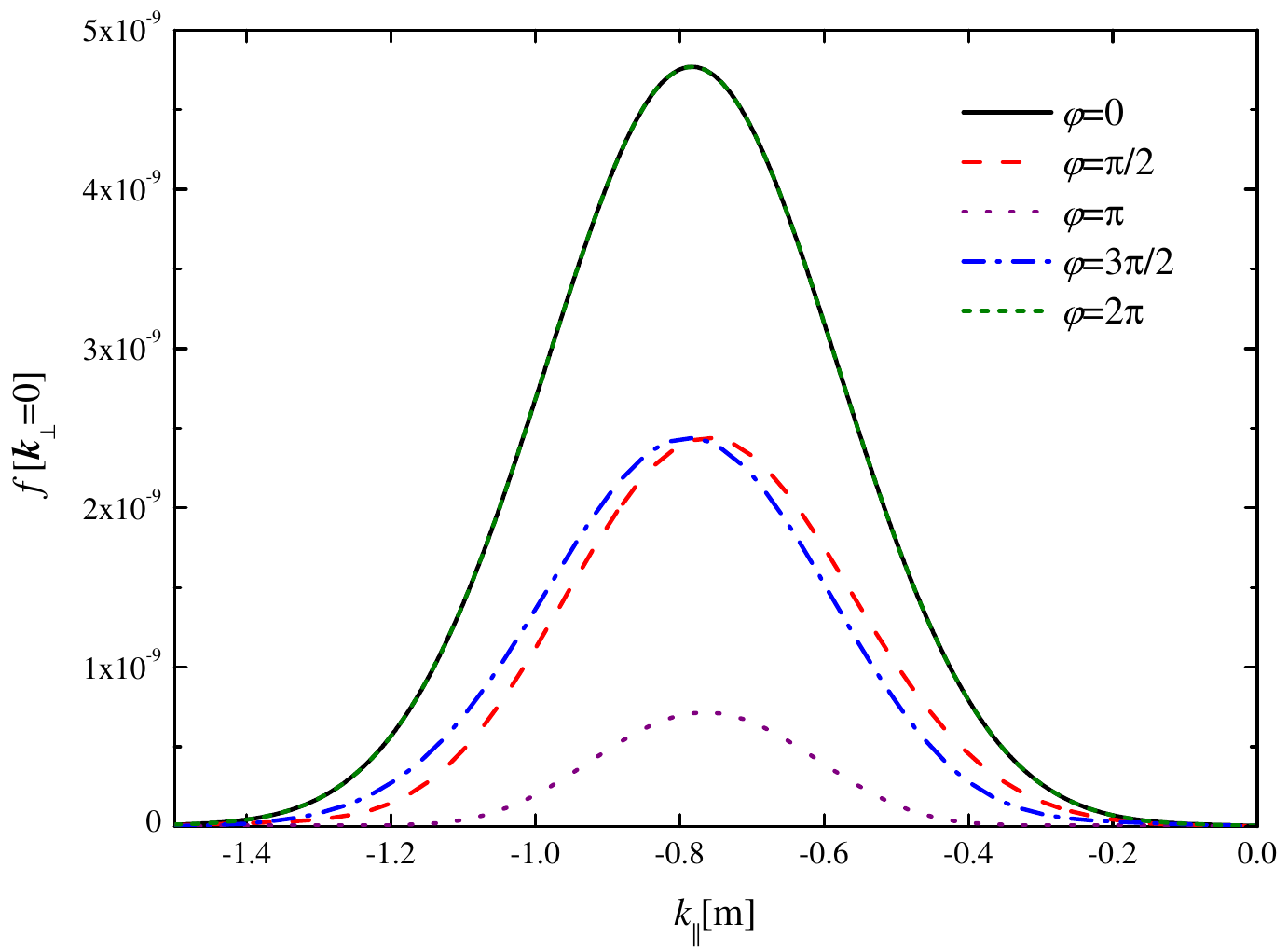}
\caption{(color online). The longitudinal momentum spectra for the combined electric field $E(t)$ with various
relative phases $\varphi$ when pulse number $N=1$ is fixed. The
electric field parameters are
$E_{1s}=0.1 E_{cr}$, $\tau_{1}=1/0.02$, $\omega_{1}=0.05$,
$E_{2w}=0.01 E_{cr}$,
$\tau_{2}=1/0.22$, $\omega_{2}=0.5$ and $T_{D}=400.3$ .}
\label{fig:6}
\end{figure}

Now consider the effect of relative phase $\varphi$ for the combined electric field $E(t)$ on the pair production.
Fig.\ref{fig:6} displays the longitudinal momentum spectra for the combined electric field $E(t)$ with various relative phase $\varphi$ when pulse number $N=1$ is fixed. Here, the relative phase $\varphi$ is chosen as $0$, $\pi/2$,
$\pi$, $3\pi/2$ and $2\pi$. It can be clearly seen that the longitudinal momentum spectra of created particles in the odd-pulse combined electric field
with various relative phase $\varphi$
is symmetric. They can also be understood from the viewpoint of
the external electric field symmetry.
For $\varphi=0$ and $\varphi=2 \pi$ the longitudinal momentum spectra
are the same from each other, while for $\pi/2$ and $3\pi/2$ they are the different. Thus, we can obtain that the cycle of the pair production process
changing with the relative phase is $2\pi$.
We think the above analysis
is universal in pair production for a external field that we considered.
Moreover, we find that the peak value of the longitudinal momentum spectra
is on the right for $\varphi=\pi/2$ and on the left for $\varphi=3 \pi/2$ comparing with that for $\varphi=0$. These results can be explained through
the configuration of the combined field. We know that the maximum value of
the combined electric field reaches at $t\textless0$ for $\varphi=\pi/2$, as a consequence the
produced particles in this electric field will be decreases for a long time.
However, when $\varphi=3 \pi/2$, the created particles can be decrease for a short time because of the maximum value of the combined electric field reaches at $t\textgreater 0$. When $\varphi=\pi$, the maximum value of the
single strong low-frequency and weak high-frequency field
composing the combined electric field both reach their maximum values at $t=0$,
but in opposite directions and the distribution of the combined field is symmetrical, which leads to the peak value of the longitudinal momentum spectrum is minimum.

\begin{figure}[htbp]\suppressfloats
\includegraphics[scale=0.5]{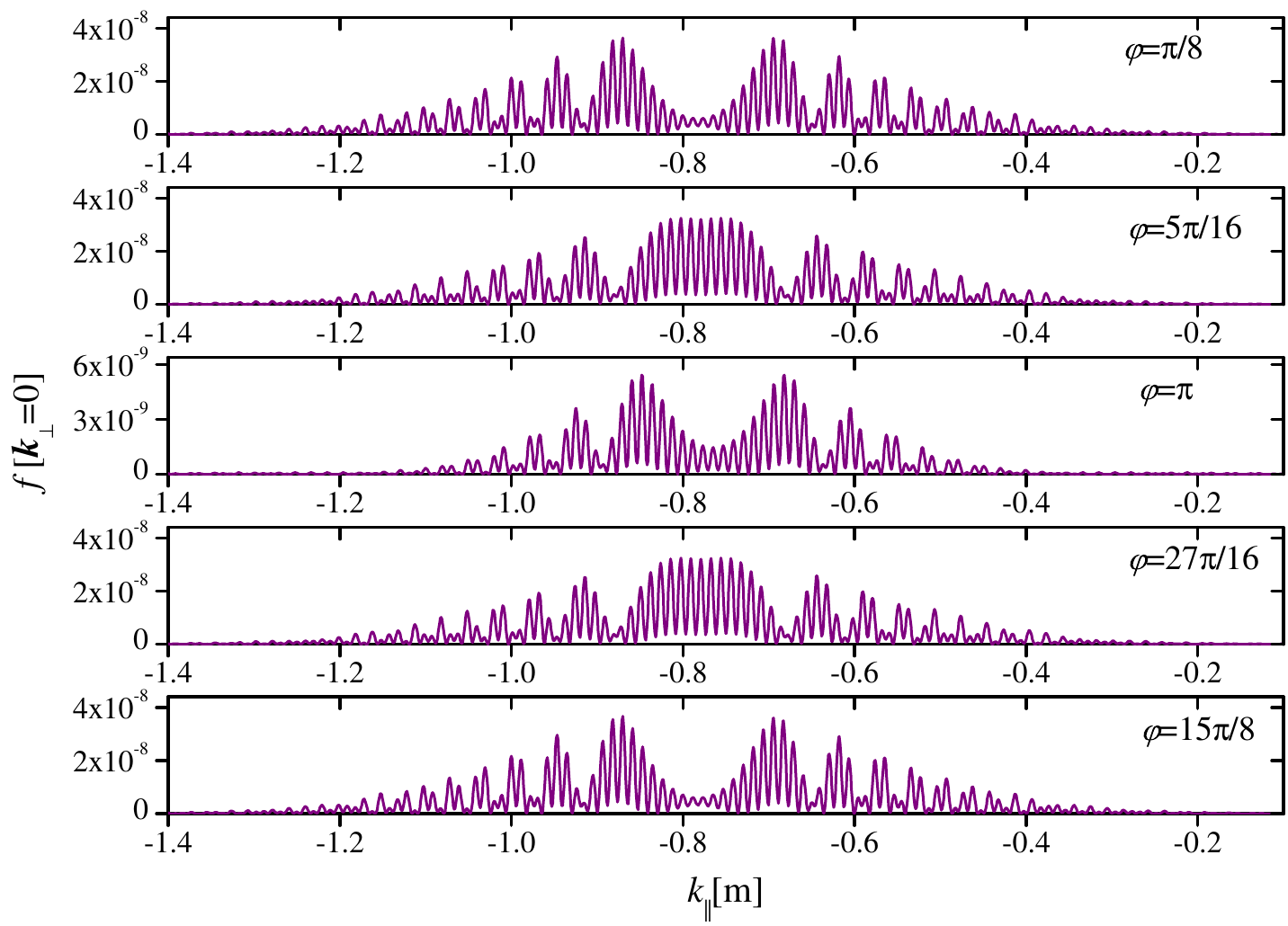}
\caption{(color online). The longitudinal momentum spectra for the combined electric field $E(t)$ with various
relative phases $\varphi$ and pulse number $N=3$. The electric field parameters are
the same as in Fig. \ref{fig:6}.}
\label{fig:7}
\end{figure}

Consider combined electric fields with pulse number $N=3$, and the longitudinal momentum spectra for the combined electric field $E(t)$ with various
relative phase $\varphi$ are sketched in Fig.\ref{fig:7}.
It is found that the momentum spectra of $\varphi=0$ and $\varphi=2 \pi$ are the same from each other at $N=1$ for combined electric field is an universal result. In addition, the momentum spectrum of produced particles in the combined field of $\varphi=0$ is discussed
in Fig.\ref{fig:1}, thus the relative phases $\varphi$ selected in the following studies are $\pi/8$,$5\pi/16$, $\pi$ ,$27\pi/16$ and $15\pi/8$, respectively.
From Fig.\ref{fig:7}, one can clearly see the symmetry of the longitudinal momentum spectra for different relative phases. This result can be also understood through the symmetry of electric field that we used.
Comparing the top and bottom panels of Fig.\ref{fig:7}, one can find that the
longitudinal momentum spectrum for $\varphi=\pi/8$ are the same as that for $\varphi=15 \pi/8$ and the momentum distribution function $\mathbf{k}_{\perp}=0$ are small nonzero value nearly around the central momentum.
When $\varphi=5\pi/16$ and $\varphi=27\pi/16$, the momentum distribution function $\mathbf{k}_{\perp}=0$ are large value nearly around the central momentum.
For the intermediate relative phase $\varphi=\pi$, the momentum spectrum is qualitatively similar to that shown in the top and bottom panels of Fig.\ref{fig:7} but the peak value is decreased significantly compared to the other cases. This result that can also be explained in terms of the field strength of the combined field.


\begin{figure}[htbp]\suppressfloats
\includegraphics[scale=0.5]{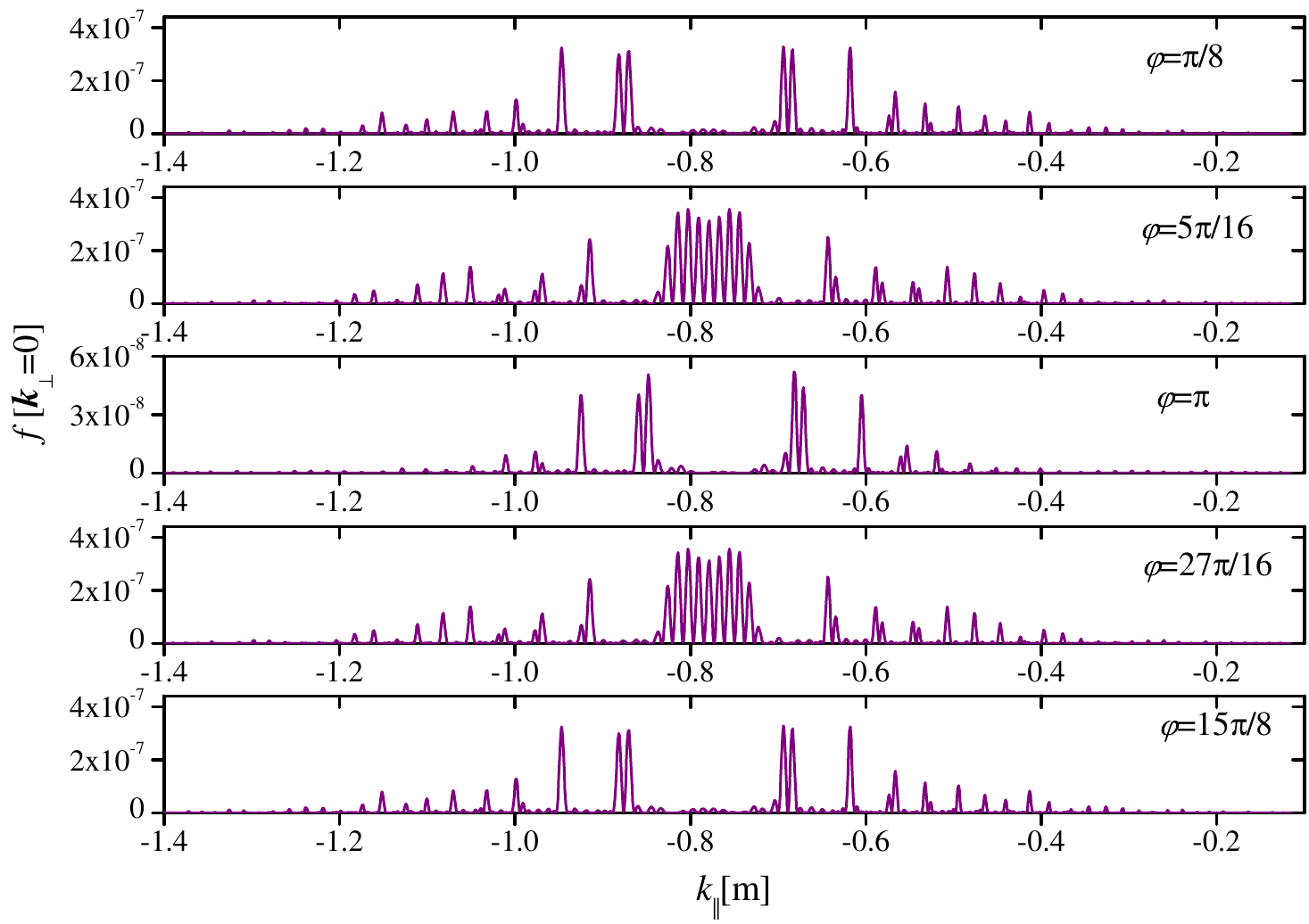}
\caption{(color online). The longitudinal momentum spectra for the combined electric field $E(t)$ with various
relative phases when $N=10$ is fixed. The field parameters are
the same as in Fig. \ref{fig:6}.}
\label{fig:8}
\end{figure}

For the large pulse number $N=10$, the longitudinal momentum spectra for $E(t)$ with various relative phases is depicted in Fig.\ref{fig:8}. In this case, the asymmetry of longitudinal momentum spectra is seen in Fig.\ref{fig:8} as strong asymmetry of the electric field, leads to such a behavior on the longitudinal momentum spectra of created particles.
The characteristic of longitudinal momentum spectrum for the relative phases $\pi/8$ and $15\pi/8$ as well as $5\pi/16$ and $27\pi/16$ that they are symmetric about $\varphi=\pi$ shown in Fig.\ref{fig:8} is the same as that for $N=3$.
However, there are some quasi-single peaks in the longitudinal momentum distribution of the created particles when the pulse number $N=10$. Accordingly, some quasi-monoenergetic particles can be obtained for the specified momentum.

\begin{figure}[htbp]\suppressfloats
\includegraphics[scale=0.5]{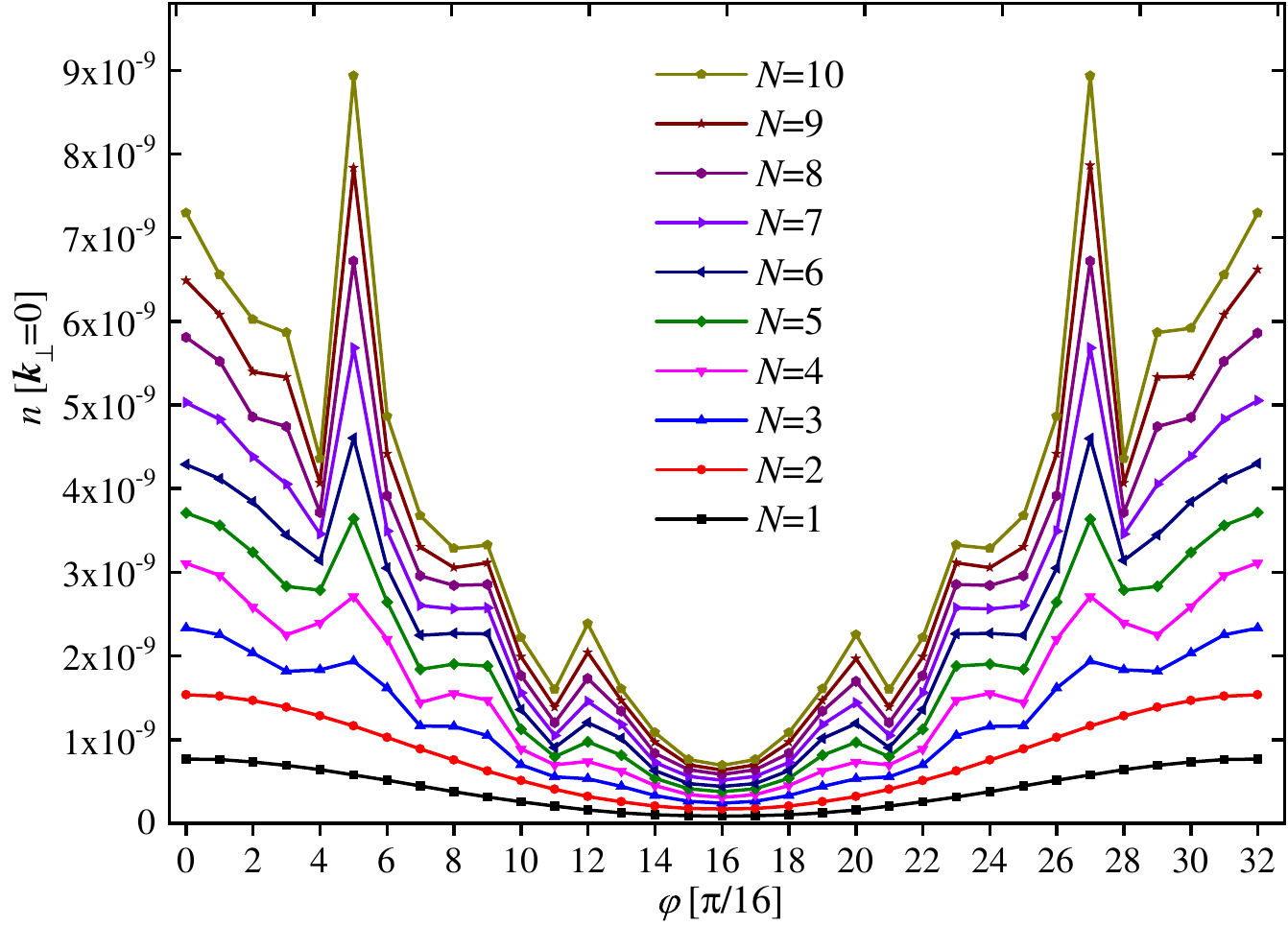}
\caption{(color online).Particle number density
$n(\mathbf{k}_{\perp}=0)$ changes with
relative phase $\varphi$ for the combined electric field $E(t)$
with different electric field pulse number $N$.
The field parameters are consistent
with those in Fig. \ref{fig:6}.}
\label{fig:9}
\end{figure}
In the case of the vanishing transverse momentum $\mathbf{k}_{\perp}=0$, the particle number density as a function of the relative phase for the combined electric field with different pulse numbers $N$ is displayed in Fig. \ref{fig:9}.
For small pulse numbers ($N\leq2$),
the number density appears minimum value at $\varphi=\pi$, while for $\varphi=0$ and $\varphi=2\pi$ it is the maximum value, which can be understood from the configuration of the combined electric field proposed in Eq. (\ref{FieldMode}). When $\varphi=\pi$, the field strengths of the single strong low-frequency and weak high-frequency field
composing the combined electric field both reach the maximum value
but have the opposite direction, so that the maximum field strength of the combined field is $E(t=0)=0.1E_{cr}-0.01E_{cr}=0.09E_{cr}$, which leads to the electron-positron pair creation become weaker than that for other relative phases of $\varphi$. However, they are maximum but have the same direction for $\varphi=0$ and $\varphi=2\pi$, that is the maximum field strength of the combined field becomes $E(t=0)=0.1E_{cr}+0.01E_{cr}=1.01E_{cr}$, thus the maximum value of particle number density is expected at $\varphi=0$ and $\varphi=2\pi$.
Moreover, we find that for $N \geq 3$, the number density reaches minimum at $\varphi=\pi$, while
for $\varphi=5/16 \pi$ and $\varphi=27/16 \pi$ it is the maximum value. In addition, there are many extreme values at $\varphi=9/16 \pi, 12/16 \pi$ and $\varphi=23/16 \pi, 20/16 \pi$.
These results indicate that the number density is sensitive to the relative phase and exhibits pronounced nonlinear structure especially for larger pulse numbers ($N\geq3$).

From Fig. \ref{fig:9}, it can also be seen that the number density
increases with the increasing pulse number $N$ for a certain relative phase $\varphi$.
Combining the above discussion we can get that for a certain pulse number $N$, the variation of the number density with the relative phase has a cycle of $2\pi$ and is symmetric about $\pi$. On the other hand, we can obtain that the relative phase has enhancement effect on the electron-positron pair production and the optimal relative phase parameters for the number density. Therefore the suitable relative phase value is very important for enhanced pair creation.

\subsection{Results for the full momentum space}\label{FM}
In this subsection, we explore the momentum spectrum and number density
of created particles for the various electric fields with nonzero transverse
momentum, i.e., full momentum space.

\begin{figure}[htbp]\suppressfloats
\includegraphics[scale=0.6]{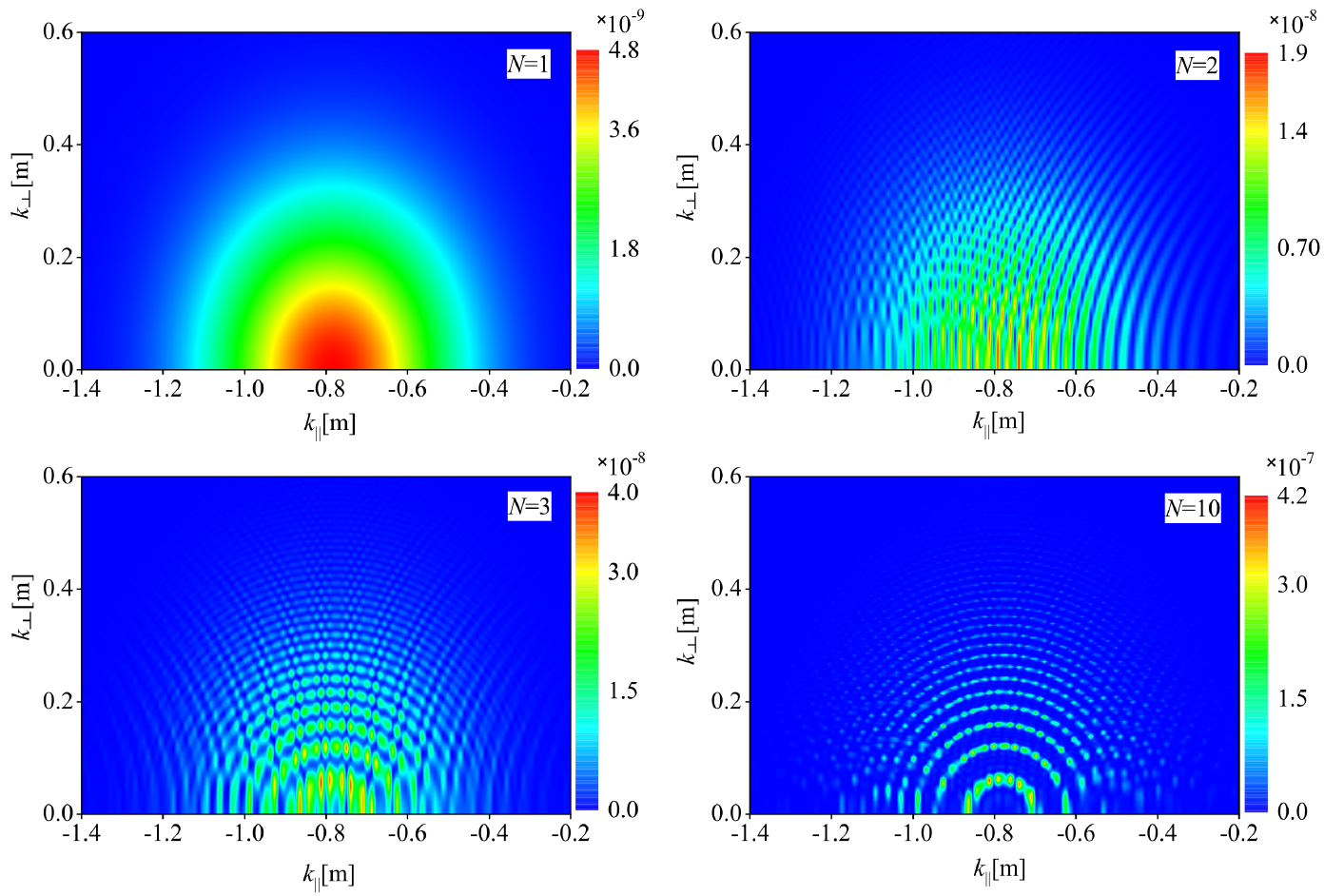}
\caption{(color online). Momentum spectra of
created particles in the combined electric field
$E(t)$ with various electric field pulse number of $N$
as $1$, $2$, $3$ and $10$. The electric field parameters are consistent
with those in Fig. \ref{fig:1}.}
\label{fig:10}
\end{figure}

\begin{figure}[htbp]\suppressfloats
\includegraphics[scale=0.6]{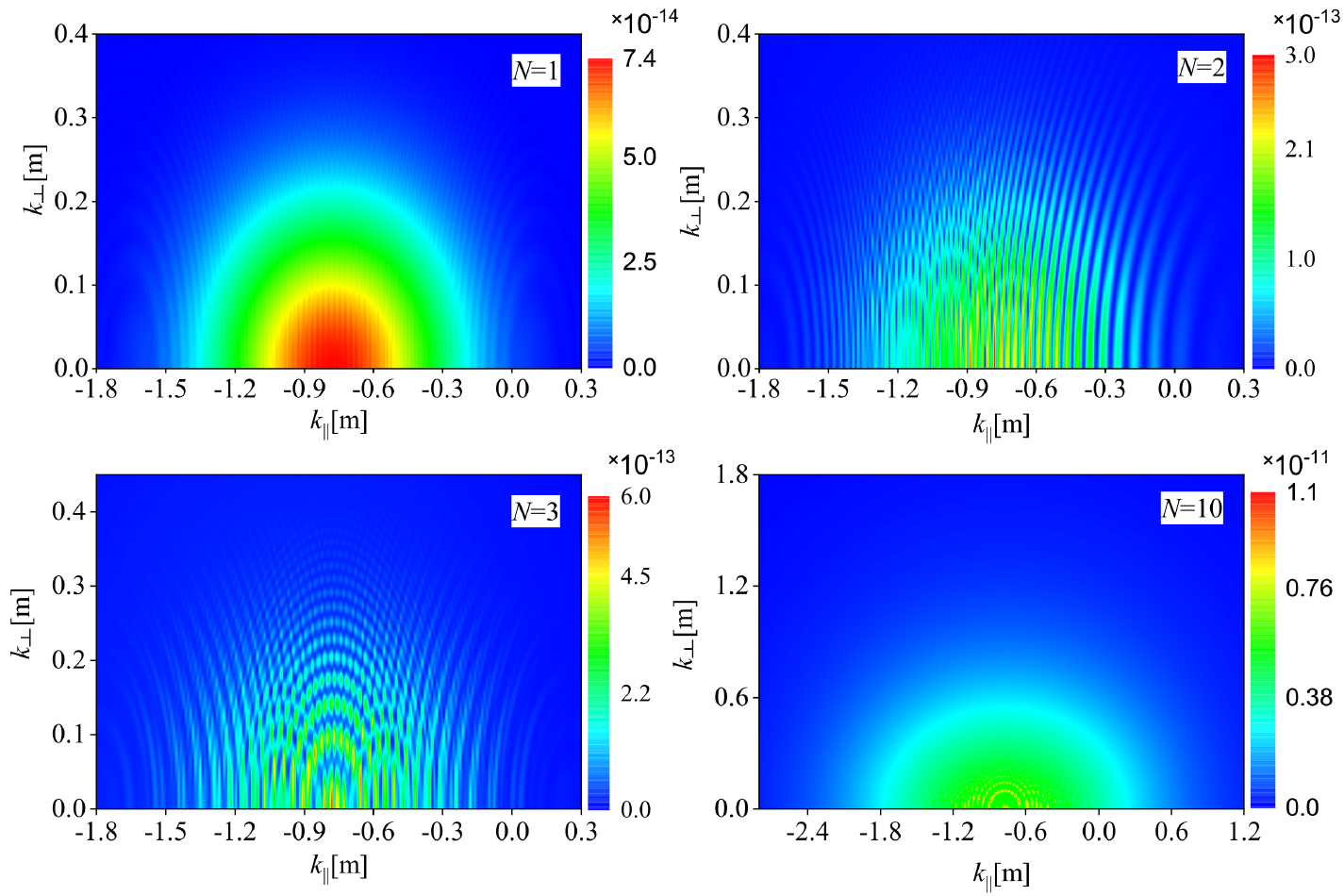}
\caption{(color online). Momentum spectra of
created particles in the low-frequency strong field pulse train $E_{1s}(t)$ with various electric field pulse number of $N$ as $1$, $2$, $3$ and $10$. The electric field parameters are consistent
with those in Fig. \ref{fig:2}.}
\label{fig:11}
\end{figure}

Let us now focus on the dependence of momentum spectrum on the pulse number for different electric field configurations. For instance, in Fig.\ref{fig:11} we display the momentum spectra of created particles in the combined electric field
$E(t)$ with various pulse numbers, i.e., $N=1$, $2$, $3$ and $10$. We can see that the momentum spectrum
shows a Gaussian-like shape for the single-pulse combined field, i.e., $N=1$, while it presents an obvious
oscillation structure for N-pulse trains electric field ($N \textgreater1$) and becomes more complex compared with that of the previous studies\cite{Li:2014psw, Akkermans:2011yn}. Also, the peak value of the momentum spectrum increases with the increase of the pulse number.
From the perspective of the scattering picture\cite{Dumlu:2010ua,Dumlu:2011rr,Nuriman:2012hn}, the momentum distribution function of created particles is approximately equal to the magnitude squared of the reflection probability equal to the square of the reflection probability  for $E\ll E_{cr}$. That is the resonance effects are determined by the positive energy $(m^2+\mathbf{k}_{\perp}^2)$ and the negative finite periodic potential $-(k_{\parallel}-eA(t))$, meanwhile it rises when some suitable value transverse momentum are given. Accordingly, the electron-positron pair production can be enhanced significantly.
In addition, the maximum value of the distribution function $f(\mathbf{k},+\infty)$ for $N$-pulse electric field does not satisfy $N^2$ times that for the single-pulse electric field compared with the result of previous studies\cite{Li:2014psw, Akkermans:2011yn}, which indicates that the pair production process is sensitive to the oscillation structure of the external field. The momentum spectra are symmetric for electric field with odd pulse numbers $N=1,3$ and asymmetric for the electric field with even pulse numbers $N=2,10$. This result occurs in the symmetry dependence of the momentum distribution function and the electric field.

Fig. \ref{fig:11} exhibits the momentum spectra of
created particles in the low-frequency strong field pulse train $E_{1s}(t)$ with various electric field pulse numbers. The corresponding pulse number $N$ is $1$, $2$, $3$ and $10$.
We can see that the interference effects still exist in the momentum spectrum for the single-pulse electric field.
In this case, we obtain the same structure of turning points as in the analysis of
Fig. \ref{fig:2}, which indicates that the momentum spectrum of the single-pulse $E_{1s}(t)$ displays interference effect is reasonable for full momentum space.
For $N \textgreater 1$, it can be seen that the oscillatory behavior of the momentum spectra becomes remarkable, meanwhile the peak value of the momentum spectrum increases significantly compared with that of the single-pulse electric field.
Furthermore, for a large pulse number $N=10$, the range of electron-positron pair production widens because the low-frequency strong field pulse train $E_{1s}(t)$ oscillation slows down and causes particle acceleration.

\begin{figure}[htbp]\suppressfloats
\includegraphics[scale=0.6]{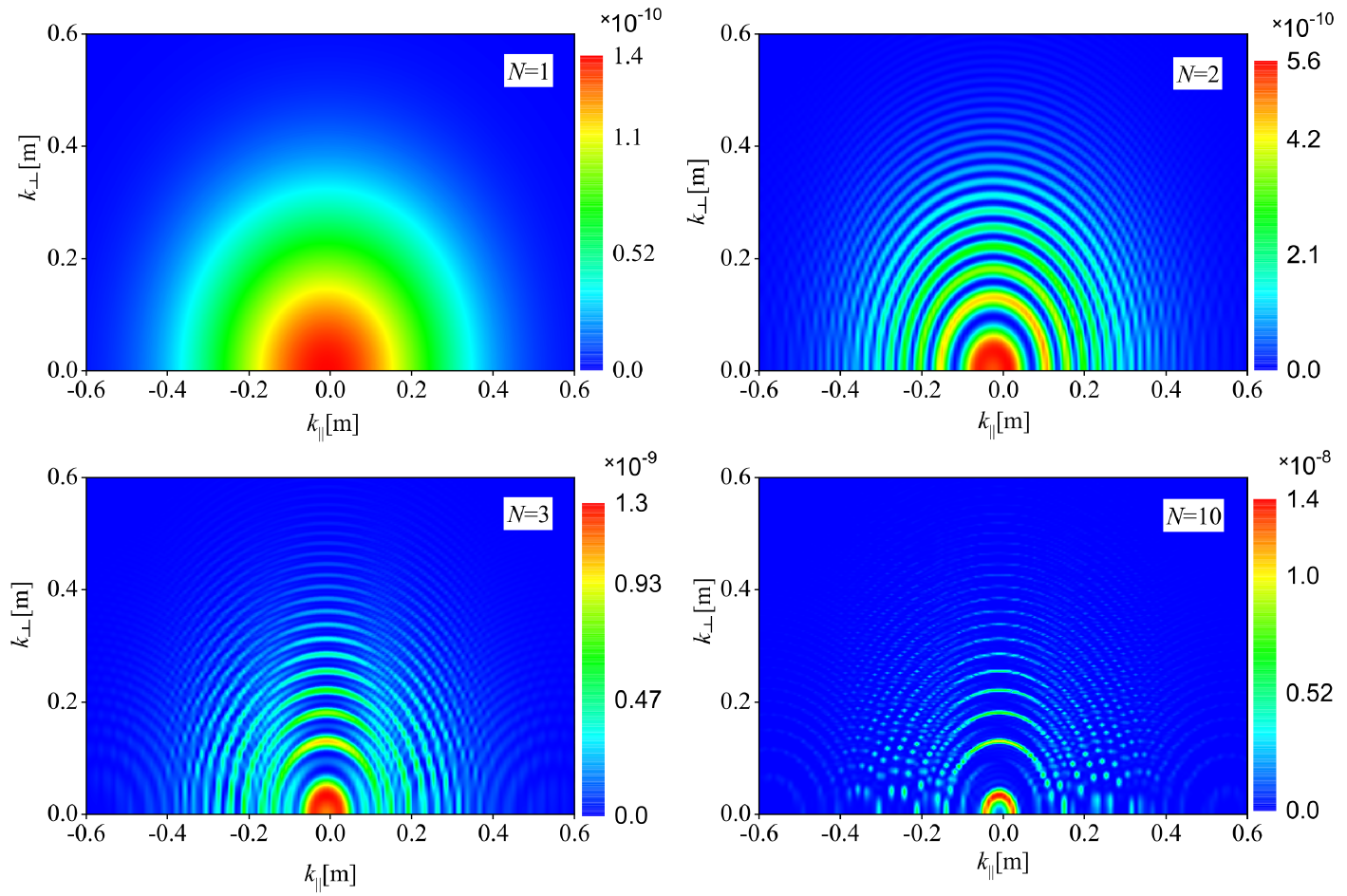}
\caption{(color online).  Momentum spectra of
created particles in the high-frequency weak field pulse train $E_{2w}(t)$ with various electric field pulse numbers of $N$ as $1$, $2$, $3$ and $10$. The electric field parameters are consistent
with those in Fig. \ref{fig:3}.}
\label{fig:12}
\end{figure}

A similar investigation is expected for the high-frequency weak field pulse train $E_{2w}(t)$ with various electric field pulse numbers, the momentum spectra of created particles are exhibited in Fig. \ref{fig:12}.
The symmetry of the momentum spectrum for different pulse numbers mentioned in the case of combined electric field $E(t)$ and the low-frequency strong field pulse train $E_{1s}(t)$ can be also observed
for low-frequency strong field pulse train $E_{2w}(t)$.
However, there are some new phenomena. From Fig. \ref{fig:12}, one can see that the range of the electron-positron pair production shrinks for the electric field $E_{2w}(t)$ with various pulse numbers.
In addition to the Gaussian-like shape momentum spectrum when the pulse number $N=1$, the ring structure appears in the momentum spectrum when $N\textgreater1$.
Combining Fig. \ref{fig:11} and Fig. \ref{fig:12}, we can clearly see that the momentum spectrum for $E_{1s}(t)$ and $E_{2w}(t)$ both display distinctly nonlinear behaviors but have obviously difference. This result indicates that the multiple-slit interference effect between non-perturbative Schwinger mechanism and the perturbative multi-photon electron-positron pair production process are different.
The deeper understand of the physical mechanism behind the phenomenon needs to study further.

\begin{figure}[htbp]\suppressfloats
\includegraphics[scale=0.5]{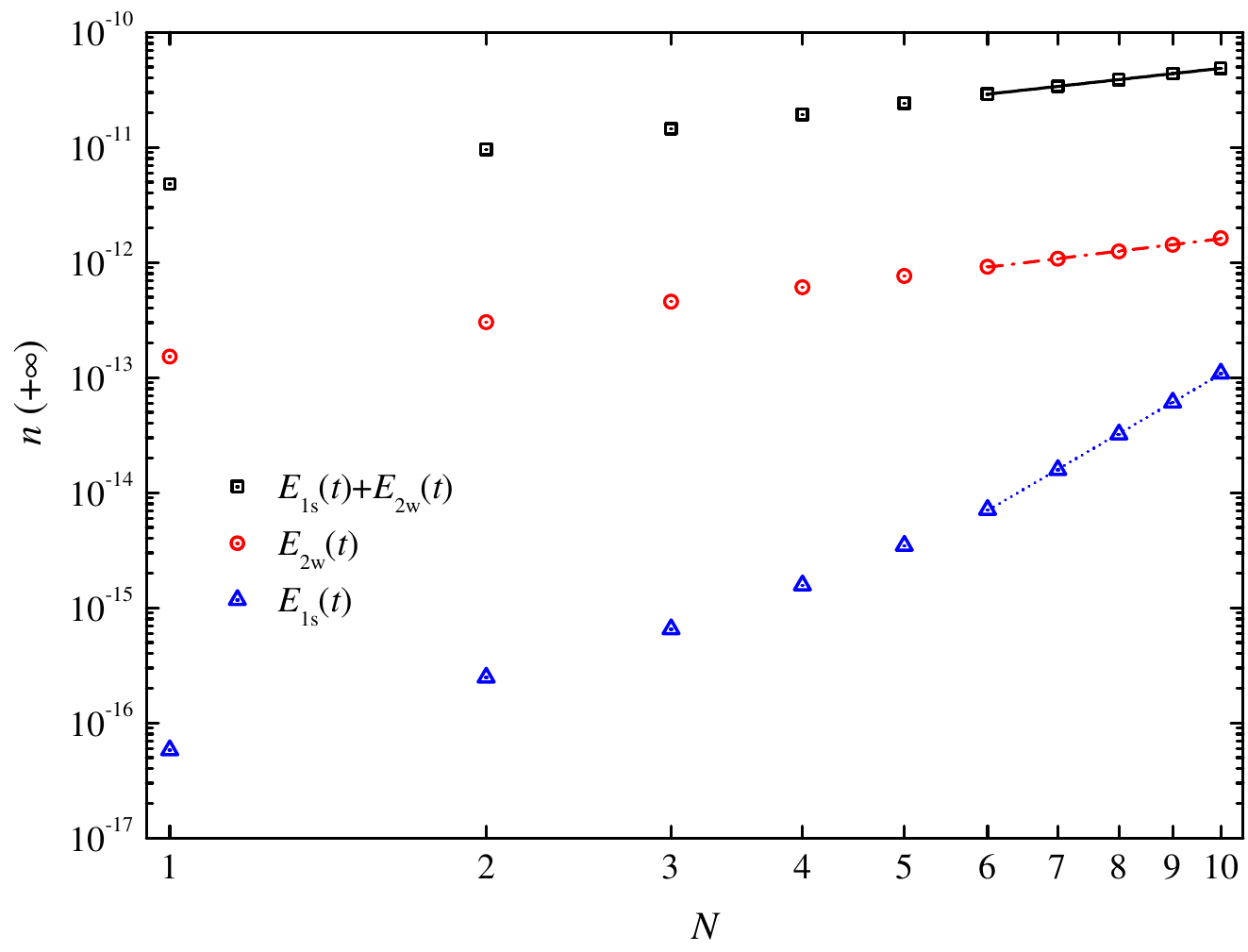}
\caption{(color online). Particles number density
$n(+\infty)$ changes with different pulse numbers $N$ for the combined electric field $E(t)$ (black squares), the electric field $E_{2w}(t)$ (red circles) and the electric field
$E_{1s}(t)$ (blue triangles), respectively. The black solid , red dash dotted and
blue dotted lines are the fitted ones for $E(t)$, $E_{2w}(t)$ and $E_{1s}(t)$
with the slopes approximate to $1.005$,$1.022$ and $5.342$. The other parameters are consistent with those in Fig. \ref{fig:1}. }
\label{fig:13}
\end{figure}

Fig. \ref{fig:13} shows the variation of the number density $n(+\infty)$ of created particles with the electric field pulse number $N$ for different electric field configurations. From Fig. \ref{fig:13}, one can see that for the three different external field configurations, the particle number density $n(+\infty)$ increases significantly with the increase of the pulse number $N$.
The black squares, red circles and blue triangles represent the results for the combined electric field $E(t)$, the electric field $E_{2w}(t)$ and electric field $E_{1s}(t)$, respectively. The black solid line, the red dashed line and the blue dotted line are the fitting lines of the corresponding results for $E(t)$, $E_{2w}(t)$  and $E_{1s}(t)$ with large pulse numbers ($N \geq 6$) and the slopes are approximately $1.005$,$1.022$ and $5.342$. Thus, we can obtain that the created particle number density and the pulse number satisfy an almost linear relationship for $E(t)$ and $E_{2w}(t)$, but a strong nonlinear relationship for $E_{1s}(t)$.
This results is different from those without oscillatory structure investigated in Ref.\cite{Li:2014psw}, indicating that the electron-positron pair production is extremely sensitive to the external field configuration.

\begin{figure}[htbp]\suppressfloats
\includegraphics[scale=0.5]{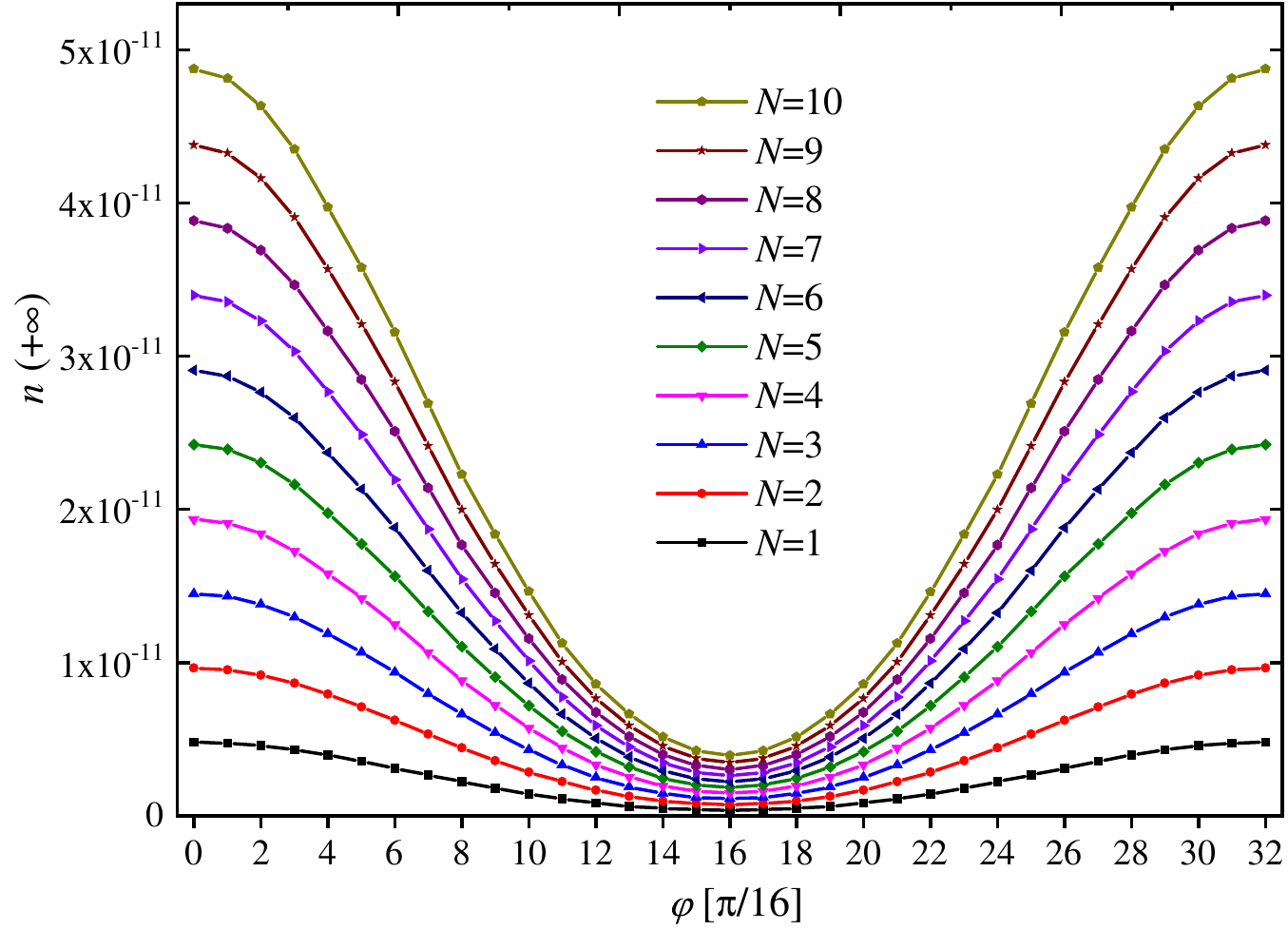}
\caption{(color online). Particle number density
$n(+\infty)$ changes with
relative phase $\varphi$ in the combined electric field $E(t)$ with different
electric field pulse number $N$. The field parameters are consistent
with those in Fig. \ref{fig:6}.}
\label{fig:14}
\end{figure}

Now we consider the number density of created particles $n(+\infty)$ changing with the relative phase $\varphi$ for the combined electric field $E(t)$ with different electric field pulse numbers $N$ in the full momentum space. Our corresponding result is
depicted in Fig. \ref{fig:14}.
First, we can see that the particle number density increases with increasing of pulse number $N$ for a given relative phase.
Specifically, when the pulse number is chosen large as $N=10$, the particle number density is increased by at least $5$ times compared with that when $N=1$.
Second, the variation trend of the particle number density  with the relative phase at different pulse numbers is the same.
The number density decreases at the relative phase $ 0\leq \varphi \leq \pi$ and increases at the relative phase $ \pi \leq \varphi \leq 2\pi$ with increasing relative phase $\varphi$. In the case of the full momentum space, we can obtain that
the number density of created particles always reach maximum at $\varphi=0$ and $\varphi=2\pi$ and minimum at $\varphi=\pi$ for the combined electric field with various pulse numbers .
From Fig. \ref{fig:14}, one can see that there is no oscillatory structure of the number density as a function of relative phase, compared to the results in the case of zero transverse momentum $\mathbf{k}_{\perp}=0$ shown in Fig. \ref{fig:9}.
It indicates that the number density of created particles is sensitive to the transverse momentum and suggests that it is worthwhile for further investigation.

\section{Summary}

In the quantum kinetic framework, we have investigated the electron-positron pair production under different alternating-sign multi-pulse trains electric fields with oscillation. The combined electric field $E(t)$, the strong but slowly varying electric field $E_{1s}(t)$ and the weak but rapidly changing electric field $E_{2w}(t)$ are considered for both the vanishing  transverse momentum space and the full momentum space.
We mainly studied the influences of the pulse number and the relative phase between the electric fields $E_{1s}(t)$ and $E_{2w}(t)$ on the momentum spectrum and the number density of created particles.
Moreover, we analyzed some nonlinear behaviors of the momentum spectra and number density of created particles from the view point of scattering potential.

For the vanishing transverse momentum space, it is found that
the longitudinal momentum spectrum become monochromatic at large pulse numbers and some suitable relative phases, which suggests that we can obtain quasi-monoenergetic particles by optimizing the pulse numbers and relative phases.
The number density increases significantly with increasing pulse number, and it is increased obviously due to the introduction of oscillatory structures in the considered background electric field.
Especially in the weak and rapidly changing electric field with different pulse numbers, the particle number density increases by more than two orders of magnitude compared to the results in the non-oscillating external fields.
Moreover, we found that the relationship between the number density and the relative phase exhibits complex nonlinear phenomenon, so that we can obtain some optimal relative phase parameters where the number density reaches maximum value.

For the full momentum space, the corresponding results show that the symmetry of the created particle momentum spectra is related to the symmetry of the external fields. In addition, it is found that the number density as a function of pulse number satisfies the power law with index $5.342$ for the strong but slowly varying electric with large pulse numbers. The variation trend of the particle number density with the relative phase for different pulse numbers is the same, which we can understood as the contribution of the transverse momentum. With the given investigation for full momentum space, the transverse momentum should be worthwhile for the further study and its physical mechanism behind the phenomenon is worthy to be studied furthermore in the future.

We investigated the electron-positron pair production in complex pulse sequences, and obtained a deeper understanding of the distinctive signatures in the momentum spectrum, but also greatly enhanced particle number density.
According, the coupling of multi-slit interference effect and dynamically assisted Schwinger mechanism for different external field configurations is still considered as an important route to increase pair production rate.
All these results can provide a significant reference for further theoretical study and a novel possibility in planning appropriate experiments. Our following work on vacuum pair production will expand to multi-pulse trains electric fields with spatial inhomogeneity and its mainly driven by the need to understand the effect for more realistic field configurations which may vary in time and space.

\newpage

\begin{acknowledgments}

\noindent
This work was supported by the National Natural Science Foundation of China (NSFC) under
Grant No. 12375240 and No. 11935008, and partially by the National Science
Foundation of GanSu Province under Grant No. 22JR5RA374.
The computation was carried out at the HSCC of the Beijing Normal University.

\end{acknowledgments}

\end{document}